\newcommand{\SeBa}{\mbox{${\sf SeBa}$}}
\newcommand{\msun}{\mbox{${\rm M}_\odot$}}
\newcommand{\Zytkow}{$\dot{\mbox{Z}}$ytkow}
\newcommand{\TZO}{$\mbox{T}\dot{\mbox{Z}}$\mbox{O}}

\documentstyle[12pt,./pasp]{article}
\def\apgt{\ {\raise-.5ex\hbox{$\buildrel>\over\sim$}}\ }
\def\aplt{\ {\raise-.5ex\hbox{$\buildrel<\over\sim$}}\ }
\begin{document}

\title{Fun for two}

\bigskip
\bigskip

\author{Simon F.\ Portegies Zwart\altaffilmark{1}}
\affil{Massachusetts Institute of Technology, Cambridge, MA 02139,
USA, Hubble Fellow}
\author{Lev R. Yungelson\altaffilmark{2}}
\affil{Institute of Astronomy of Russian Academy of Sciences, Moscow,
Russia}
\author{Gijs Nelemans\altaffilmark{3}}
\affil{Sterrenkundig Institute {\em Anton Pannekoek}, Amsterdam, The Netherlands}

\bigskip
\bigskip


\begin{abstract}

We performed populations synthesis calculations of single stars and
binaries and show that binary evolution is extremely important for
Galactic astronomy. We review several binary evolution models and
conclude that they give quite different results.  These differences
can be understood from the assumptions related to how mass is
transfered in the binary systems.  Most important are 1) the fraction
of mass that is accreted by the companion star during mass transfer,
2) the amount of specific angular momentum which is carried away with
the mass that leaves the binary system.

\end{abstract}
\section{Introduction}

Binaries are characterized by ``the union of two stars, that are
formed together in one system, by the laws of attraction'' (Herschel
1802).  They from the basic building blocks of the Milky Way as
galaxies are the building blocks of the Universe.  In the absence of
binaries many astrophysical phenomena would not exist and the Galaxy
would look completely different over the entire spectral range.  A
considerable fraction of the astrophysical community would be
unemployed, Doppler (1842) would not have written his famous paper,
Herschel's nearest neighbor distribution of field stars would look
different. Even life as we know it would not have evolved in the
Universe as Type Ia supernovae, which enrich the interstellar medium
with elements required to enable life, would not occur.

\section{The Galaxy with single stars}

Let's assume that the Galaxy contains only single stars and that the
evolution of a single star passes through three stages; main-sequence
(ms), giant (gs) and remnant, which again we subdivide into white
dwarf (wd) and neutron star (ns), black holes are neglected here.
Table\,\ref{Tab:single} presents the distribution of stars over these
subtypes at different times during the lifetime of the Galaxy if all
stars were born at the same time (columns 2--5) and if the star
formation was constant for the last 10\,Gyr (column 6).  We assumed
that the initial mass function is given by the distribution proposed
by Scalo (1986) between 0.1\,\msun and 100\,\msun. Calculations were
performed with the \SeBa\ population synthesis code (Portegies Zwart
\& Verbunt 1996, see the starlab software tool set {\tt
http://www.sns.ias.edu/$\sim$starlab}).

\begin{table}[htbp!]
\caption{Stellar types and total mass as a function of
time.  Calculation are performed with $10^5$ stars, but the numbers
are rescaled to 100 stars.  The first column gives the stellar type
followed by the normalized number of stars of that type at zero age,
100\,Myr, 1\,Gyr and 10\,Gyr after formation.  The last column gives
the stellar population if the star formation rate was constant over
10\,Gyr.  The bottom line gives total mass in \msun.  }
\medskip
\begin{tabular}{lr|rrrrrr} \hline
time $[$Myr$]$&      0&   100&  1000& 10000& 0--10000 \\ \hline		      
ms            &    100& 99.03& 94.98& 86.24& 90.14  \\
gs            &      0&  0.22&  0.52&  0.85&  0.80 \\
wd            &      0&  0.41&  4.16& 12.57&  8.72 \\
ns            &      0&  0.34&  0.34&  0.34&  0.34 \\
mass $[\msun]$ &  61.6&  55.3&  46.8& 40.1& 42.9 \\ \hline
\end{tabular}
\label{Tab:single} 
\end{table} 

At all times the stellar population is dominated by main sequence
stars (of which the majority lives longer than 10\,Gyr), followed by
white dwarfs and giants. The $\sim 14$\% of the stars which evolve
reduces the total mass with $\sim 35$\%.  The population of neutron
stars builds up within a couple of 10\,Myrs and remains constant at
later times.

\section{The Galaxy with binaries}

If we fill the Galaxy with binaries things become more interesting and
considerably more complicated (see Table\,\ref{Tab:binary}).

Except for the initial mass function we now have to select the mass of
the secondary star, the orbital period and the ellipticity of the
binary system. In our numerical experiment these were all selected
following model $A$ of Portegies Zwart \& Verbunt (1996). The initial
conditions are representative for the G-dwarfs in the solar
neighborhood (Duquennoy \& Mayor 1991), which are well established.
The orbital elements and masses of the two stars for binaries with
other spectral types are still ill known and recent work is painfully
sparse.

Instead of evolving single stars we now have to evolve two stars
synchronously. And at the same time account for the effects of their
evolution on the orbital parameters of the binary system.  Furthermore
the evolution of each star may be affected by its companions'
evolution, i.e.\, through tidal effects and mass transfer.

Table\,\ref{Tab:binary} shows the distribution of binaries over the
various subtypes at several moments in time (as in
Table\,\ref{Tab:single}). The single stars originate from binaries
which are broken up or have coalesced.  We keep the same simple
subtypes as before and write a binary as the two stars enclosed with
parenthesis following the notation introduced by Portegies Zwart \&
Verbunt (1996).  The number of possible outcomes is much larger than
for the evolution of a population of single stars.

The total mass in binaries is slightly higher than 1.5 times the total
mass in single stars (see Table\,\ref{Tab:single}), as one would expect
from a flat mass ratio distribution.  This is becasue we limit the
masses of the primary and secondary stars to $>0.1$\,\msun.

\begin{table}[htbp!]
\caption{Stellar types from population synthesis of $10^5$
binaries. All stars are born in binaries and they are evolved in time
with \SeBa.  For binaries the two stellar types are enclosed by
parenthesis, a bracket indicates that the star is transferring mass to
its companion.  
}
\medskip
\begin{tabular}{l|rrrrr} \hline
time $[$Myr$]$& 0&    100&     1000&    10000 & 0 -- 10000 \\ \hline
ms         &    0&   0.07&     0.02&     0.01 &      0.02 \\
gs         &    0&   0.03&     0.10&     0.11 &      0.08 \\
wd         &    0&   0.09&     0.76&     2.68 &      1.90 \\
ns         &    0&   0.45&     0.46&     0.46 &      0.51 \\  \hline 
(ms, ms)   &  100&  98.60&    94.47&    84.27 &     89.07 \\  
(ms, gs)   &    0&   0.18&     0.47&     0.77 &      0.63 \\
(ms, wd)   &    0&   0.15&     1.80&     5.97 &      4.01 \\
(ms, ns)   &    0&   0.01&     0.00&     0.00 &      0.00 \\
(gs, gs)   &    0&   0.01&     0.01&     0.02 &      0.01 \\
(gs, wd)   &    0&   0.04&     0.16&     0.27 &      0.23 \\
(gs, ns)   &    0&   0.00&     0.00&     0.00 &      0.00 \\
(wd, wd)   &    0&   0.03&     1.04&     3.72 &      2.47 \\
(wd, ns)   &    0&   0.02&     0.02&     0.01 &      0.02 \\
(ns, ns)   &    0&   0.00&     0.00&     0.00 &      0.00 \\ \hline
$[$ms, ms$]$&   0&   0.51&     0.71&     1.41 &      0.96 \\
$[$gs, ms)   &    0&  0.00&     0.02&     0.01 &      0.01 \\
$[$ms, wd)   &    0&  0.00&     0.00&     0.06 &      0.03 \\
mass $[\msun]$&96.9&  87.8&     76.3&     64.9 &      69.6 \\ \hline
\end{tabular}
\label{Tab:binary} 
\end{table}

As expected, main-sequence binaries are most common but at later age a
considerable fraction of binaries contain at least one white dwarf.
Note that the fraction of single stars produced from the evolution of
binaries is small and the majority of these are white dwarfs. Most of
the single white dwarfs are the result of a merger in the common
envelope phase after the formation of the first white dwarf.

The simple representation used here is insufficient to describe the
evolution of binary stars in detail. It neglects interesting
information about the distributions of masses, mass ratios, orbital
period and eccentricities and it lacks detailed information to, for
example, distinguish blue stragglers from main sequence stars or
identify low mass carbon-oxygen white dwarfs (see for example Nelemans
et al.\, 2000).  It however, shows that the possible outcomes of the
evolution of a population of primordial binaries is vastly larger than
for single stars.

\section{Binary evolution}

There are many binary population synthesis programs available which
claim to be able to evolve any binary in time. This industry was
started in the early eighties by Kornilov \& Lipunov (1983), Iben \&
Tutukov (1984) followed by Dewey \& Cordes (1987) and continues to the
present time.

We will show an evolutionary sequence produced by three of these
programs together with the fully conservative case which is easily
computed by hand.  We can not present the evolution of more scenarios
or those produced by other models because most publications do not
provide sufficient information to follow the complete evolution of a
particular binary system.  In Table\,\ref{Tab:evolution} we show one
evolutionary track of one of the rare comparisons which can be made.

\begin{table}[htbp!]
\caption{Various stages (first column) of the evolution of a close
binary with massive stars; (A): birth $[t=0]$, (B): start $1^{\rm st}$
Roche-lobe contact $[t\simeq14.2$\,Myr$]$, (C): before $1^{\rm st}$
supernova $[t\simeq16.1$\,Myr$]$, (D): after 1$^{\rm st}$ supernova,
(E): start 2$^{\rm nd}$ Roche-lobe contact $[t\simeq19.9$\, Myr$]$,
(F): before $2^{\rm nd}$ supernova $[t\simeq20.8$\,Myr$]$, (G): after
$2^{\rm nd}$ supernova (not present in the LPP96 case).  Masses ($M$
and $m$) are in solar units, orbital period ($P$) in days.  The
various evolutionary models are: fully conservative (indicated by
Conservative), Tutukov \& Yungelson (1993, TY93), Portegies Zwart \&
Verbunt (1996,
\SeBa, see also {\tt http://ww.sns.ias.edu/$\sim$starlab}) and
Lipunov, Postnov
\& Prokhorov (1996, LPP96, see {\tt
http://xray.sai.msu.ru/sciwork/scenario.html}).  }
\medskip
\begin{tabular}{l|rcl|rcl|rcl|rcl} \hline
Stage	   &\multicolumn{3}{c}{Conservative} &\multicolumn{3}{c}{TY93}
	   &\multicolumn{3}{c}{\SeBa} &\multicolumn{3}{c}{LPP96} \\ 
	   & $M$  & $m$ & $P$  & $M$  & $m$ & $P$ 
	   & $M$  & $m$ & $P$ & $M$  & $m$ & $P$ \\
A:         & 13.1 & 9.8 & 39.2 &13.1 & 9.8 & 39.2
	   & 13.1 & 9.8 & 39.2 &13.1 & 9.8 & 39.2 \\
B:         & 13.1 & 9.8 & 39.2 & 13.1 & 9.8 & 39.2
           & 12.7 & 9.8 & 40.7 & 12.2 & 9.6 & 43.2 \\ 
C:         & 3.0 & 19.9 & 390 & 3.3 & 15.4 & 20.3
	   & 3.7 & 18.7 & 204 & 3.7 & 10.0 & 301 \\ 
D:         & 1.4 & 19.9 & 411 & 1.4 & 15.4 & 30.7
           & 1.3 & 18.7 & 259 & 1.4 & 10.0 & 241 \\
E:         & 1.4 & 19.9 & 411 & 1.4 & 15.4 & 24.6
           & 1.3 & 17.6 & 263 & 1.4 &  9.9 & 181 \\
F:         & 1.4 & 5.1 & 9.7 & 1.4 & 4.2 & 0.1
	   & 1.3 & 3.4 & 0.2 & 4.3 & \multicolumn{2}{c}{\TZO} \\
G:         & \multicolumn{3}{c|}{dissociated} & 1.4 & 1.4 & 2.2
	   & 1.3 & 1.3 & 2.4 & 4.3 & \multicolumn{2}{c}{black hole} \\
\hline
\end{tabular}
\label{Tab:evolution} 
\end{table}

Table\,\ref{Tab:evolution} shows that various groups obtain quite
different results from identical initial conditions (see also Verbunt
1996, who performed a similar comparison between two models).  The
conservative case is most radically different, indicating that all
groups agree that the evolution of such a binary should proceed rather
inconservative.  The other extreme in this example is provided by the
Scenario Machine of Lipunov et al.\, (1996) in which case the
accreting star hardly gains any mass but most mass is ejected from the
binary system (last three columns). The large period after mass
transfer indicates that little angular momentum is carried with the
lost material. (The change in orbital period can be reconstructed
assuming that mass lost from the binary carries 0.71 times the
specific angular momentum of the binary system, which is about the
specific angular momentum of the accreting star at the onset of the
mass transfer.) This model results in a single Thorne-\Zytkow\, object
(a giant with a neutron star core) after the second phase of mass
transfer.

The two models in the middle of Table\,\ref{Tab:evolution} (Tutukov \&
Yungelson and \SeBa) both lead to a neutron star binary.  In these
cases the intermediate stages of the binaries, however, are quite
different; the binary in the TY93 model remains rather close, where
\SeBa\, results in a much wider intermediate state. 
So in the TY93 case, more mass is lost carrying, on average, more
angular momentum.  In \SeBa\, mass leaves the binary system with 3
times the specific angular momentum of the binary; applying this
prescription to the model of TY93 we find that $\sim 5.3$ times the
angular momentum of the binary system is lost per unit mass. Note
however, that TY93 use a completely different treatment of
non-conservative mass transfer.  The differences in the treatement of
the common envelope in the second phase of mass transfer makes that
the final binaries computed with the TY93 and \SeBa\, models are very
similar.

\subsection{Why are the models so different}

The differences between the calculations presented in
Table\,\ref{Tab:evolution} are rather big but can be brought back to a
few assumptions about the mass transfer.  These assumptions determine
1) the fraction of mass that is accreted by the companion star during
mass transfer, 2) the amount of specific angular momentum which is
carried away with the mass that leaves the binary system.

Even bigger differences are expected from introducing new physics in
the models, which may lead to unexplored channels for the formation of
various types of binaries or to completely new classes of objects.  An
example is the model in which a neutron star in a common envelope
accretes a significant fraction of this envelope.  This causes the
neutron star to grow in mass until it exceeds the stability limit and
collapses into a low-mass black hole (Chevalier \& Kirsner
1979). Bethe \& Brown (1998) used this new understanding to calculate
the number of neutron star binaries in the Galaxy with an analytic
model.  The computer powered population calculations of Portegies
Zwart \& Yungelson (1998) gave identical numbers for the models with
similar assumptions (see their model $H$).  The agreement between the
two completely different techniques indicates that the uncertainties
in binary population synthesis are mainly caused by differences in the
assumption about the underlying physics and in a lesser extent in the
proper choice of various key parameters.  By lack of a proper
understanding of some of the background physics these parameters are,
for now, to be adjusted such that the observed binaries can be
explained.

\bigskip\noindent{\bf Acknowledgments} This work
was supported by NASA through Hubble Fellowship grant HF-01112.01-98A
awarded (to SPZ) by the Space Telescope Science Institute, which is
operated by the Association of Universities for Research in Astronomy,
Inc., by Spinoza Grant 08-0 to E.P.J. van den Heuvel and by RFBR grant
99-02-16037.

\end{document}